\documentclass[%
 reprint,
superscriptaddress,
nofootinbib,
 amsmath,amssymb,
pra,
]{revtex4-2}

\usepackage{graphicx}% Include figure files
\graphicspath{{figures/}}  % or {images/}
\usepackage{dcolumn}% Align table columns on decimal point
\usepackage{lipsum}
\usepackage{multirow}
\usepackage{upgreek}
\usepackage[uncertainty-mode=separate]{siunitx}
\usepackage{booktabs}
\usepackage{ragged2e}
\usepackage{bm}% bold math
\usepackage[version=4]{mhchem}
\usepackage{threeparttable}
\usepackage[hidelinks=true]{hyperref}% add hypertext capabilities
\usepackage[capitalise, nameinlink]{cleveref}

\crefname{section}{Sec.}{Sec.}
\Crefname{section}{Sec.}{Sec.}
\newcommand{\um}{\unit{\upmu\meter}}

\begin{document}

\title{Post-fabrication trimming of a 1024-pixel, single-photon counting, Microwave Kinetic Inductance Detector array with a pixel pitch of 150 micron}

\author{Wilbert Ras-Vinke}%
    \email[Contact author at ]{w.ras@sron.nl}
    \affiliation{Space Research Organisation Netherlands (SRON), Niels Bohrweg 4, Leiden 2333 CA, Netherlands}
    \affiliation{Department of Microelectronics, Delft University of Technology, Mekelweg 4, Delft 2628 CD, Netherlands}
\author{Hessel Schulte}
    \affiliation{Department of Microelectronics, Delft University of Technology, Mekelweg 4, Delft 2628 CD, Netherlands}
\author{David J. Thoen}%
    \affiliation{Space Research Organisation Netherlands (SRON), Niels Bohrweg 4, Leiden 2333 CA, Netherlands}
    \affiliation{Department of Microelectronics, Delft University of Technology, Mekelweg 4, Delft 2628 CD, Netherlands}
\author{Kevin Kouwenhoven}%
    \affiliation{Space Research Organisation Netherlands (SRON), Niels Bohrweg 4, Leiden 2333 CA, Netherlands}
    \affiliation{Department of Microelectronics, Delft University of Technology, Mekelweg 4, Delft 2628 CD, Netherlands}
\author{Steven A.H. de Rooij}
    \affiliation{Space Research Organisation Netherlands (SRON), Niels Bohrweg 4, Leiden 2333 CA, Netherlands}
    \affiliation{Department of Microelectronics, Delft University of Technology, Mekelweg 4, Delft 2628 CD, Netherlands}
\author{Tonny A.H.M. Coppens}
    \affiliation{Space Research Organisation Netherlands (SRON), Niels Bohrweg 4, Leiden 2333 CA, Netherlands}
\author{Jochem J.A. Baselmans}%
    \affiliation{Space Research Organisation Netherlands (SRON), Niels Bohrweg 4, Leiden 2333 CA, Netherlands}
    \affiliation{Department of Microelectronics, Delft University of Technology, Mekelweg 4, Delft 2628 CD, Netherlands}
    \affiliation{Physikalisches Institut, Universität zu Köln, Zülpicher Straße 77, 50937 Cologne, Germany }
\author{Pieter J. de Visser}%
    \affiliation{Space Research Organisation Netherlands (SRON), Niels Bohrweg 4, Leiden 2333 CA, Netherlands}
    \affiliation{Department of Microelectronics, Delft University of Technology, Mekelweg 4, Delft 2628 CD, Netherlands}
    
\date{\today}

\begin{abstract}
Microwave Kinetic Inductance Detectors are superconducting resonators capable of single-photon counting with energy resolving capability at visible and near-infrared wavelengths. Their zero read noise, extremely low dark counts, and compatibility with frequency-division multiplexing make them well suited for large imaging arrays. The detector yield of these arrays is often limited by the fabrication-induced frequency scatter leading to resonator collisions. We find that the tight packing of the pixels can add significant frequency scatter as well. We developed a post-fabrication correction method that is suitable for tightly packed arrays with a pixel pitch of \qty{150}{\um}. We demonstrate this method on a \num{1024}-pixel array multiplexed on a single octave of readout bandwidth and show a reduction in frequency scatter from \num{1.1e-2} to \num{3.5e-4} and an increase in yield from \qty{76}{\percent} to \qty{94}{\percent}.

\end{abstract}

\maketitle

\section{Introduction}
\label{sec:introduction}
Microwave Kinetic Inductance Detectors (MKIDs) are superconducting detectors that can count single photons and resolve their energy at visible and near-infrared wavelengths \cite{mazinARCONS2024Pixel2013}. They have microsecond timing resolution, no read noise and extremely low dark count rates \cite{walterMKIDExoplanetCamera2020, swimmerCharacterizingDarkCount2023}. MKIDs will help answer some of human's fundamental questions about life in the universe by enabling simultaneous imaging and spectroscopy of earth-like exoplanets on future space-based observatories \cite{steigerSimulatedPerformanceEnergyresolving2024, howeScientificImpactNoiseless2024a, bryanKIDDetectorReadout2025} as well as by improving high contrast imaging efforts from ground by speckle suppression \cite{meekerDARKNESSMicrowaveKinetic2018, walterMKIDExoplanetCamera2020} and multi-wavelength wavefront sensing \cite{darcisAddingColourZernike2025a, darcis2026inprep, magniezPolychromaticPyramidWavefront2024b, magniezPolychromaticPyramidWavefront2026a}. Other applications are to use MKIDs as `order resolvers' in single object spectroscopy where they can replace the complex cross dispersion optics \cite{obrienKIDSpecMKIDBasedMediumResolution2020} or as a photon-limited hyperspectral microscope in medical applications \cite{shawMKIDMicroscopes2023}.

The detection principle is based on the breaking of Cooper pairs by visible or near-infrared photons in a highly inductive superconducting film embedded in a microwave resonator \cite{dayBroadbandSuperconductingDetector2003, mazinARCONS2024Pixel2013}. A single photon breaks thousands of Cooper pairs, depending on its energy, shifting the resonator frequency. This shift is measured through changes in the phase and amplitude of a microwave probe signal. 
Because each resonator can be assigned a unique resonance frequency, MKIDs inherently support frequency-division multiplexing (FDM), allowing many detectors to share a single readout line. High multiplexing factors---the number of MKIDs coupled to a single readout line---reduce the mass, power, and cost of readout electronics, which is of particularly importance for space-based instruments. 
Several MKID instruments have seen first light in the last decade \cite{reyesAMKIDLargeKIDbased2025, calvoNIKA2InstrumentDualBand2016, karatsuDESHIMA202004002026}, the largest being a near-infrared single-photon-counting camera with over 20 kilo-pixels with a multiplexing factor of 2000 on 1 octave of readout bandwidth \cite{walterMKIDExoplanetCamera2020}. Larger MKID instruments are in development that have 38 kilo-pixels and a multiplexing factor of 4000 on 2 octaves of bandwidth \cite{huberCCATCharacterizationFirst2026a}.

However, the multiplexing factors of these MKID devices are generally limiting the yield. The yield---the fraction of usable pixels---is often limited by insufficient frequency spacing between resonators, or collisions, which occur more often for higher multiplexing factors. Collided resonators cannot be read out independently and must therefore be masked out, reducing the yield.
The yield can then be increased by giving in on the number of readout lines \cite{calvoNIKA2InstrumentDualBand2016} or readout bandwidth \cite{huberCCATCharacterizationFirst2026a}, but can also be increased more fundamentally in (post-)fabrication. Resonator collisions are generally attributed to fabrication-induced variations in resonator geometry and material properties \cite{shuUnderstandingMinimizingResonance2021, albertSpatialMappingKilopixel2024, middletonCCATLEDMapping2024}, so yield can be increased by improving the initial fabrication---like moving from contact to electron-beam lithography \cite{reyesAMKIDLargeKIDbased2025}---or by trimming the resonators in a post-fabrication step \cite{liuSuperconductingMicroresonatorArrays2017, shuIncreasedMultiplexingSuperconducting2018, shuUnderstandingMinimizingResonance2021}. 

In this work we study MKID arrays operating at visible/near-infrared wavelengths, which are significantly smaller and more densely packed than far-infrared/(sub-)millimeter MKID arrays. We find that the electron-beam lithography process remains a limiting factor to the yield. Furthermore, we find that the tight packing of our pixels significantly increases the frequency scatter of the resonances, increasing the demand for a post-fabrication correction method. However, the methods that make use of laser scanning \cite{liuCryogenicLEDPixeltofrequency2017} or contact lithography \cite{shuIncreasedMultiplexingSuperconducting2018, shuUnderstandingMinimizingResonance2021} are not applicable as these lack the required spatial resolution. 
We demonstrate a very accurate post-fabrication correction method based upon electron beam lithography, which we successfully apply on a 1024-pixel array. We show that we can reduce the overall frequency scatter of the resonators---from fabrication-induced variations and the tight pixel pitch---with over a factor 30.
% \clearpage

\section{Methods}
\label{sec:methods}
\subsection{Yield and frequency scatter}
\label{sec:yield and frequency scatter}
The yield of an MKID array is defined as the total number of pixels that can be used efficiently to perform the type of detection for which the array is designed. It is affected in three ways. Firstly, fabrication can fail, resulting in defective pixels. Secondly, resonators can collide in frequency, resulting in signal mixing between these pixels. Lastly, pixels that fail to meet certain performance criteria--- like quantum efficiency, resolving power or dynamic range---cannot be used efficiently for signal detection. In this work we discuss the yield considering fabrication and resonator collisions only. We will now give a quantitative description of resonator collisions and yield.

Suppose we have an array with $N$ resonators with designed resonance values $F_\mathrm{D}[n]$ and measured values $F_\mathrm{M}[n]$ for $n=0, 1, \dots,N-1$. Two neighboring resonances with $F_\mathrm{M}^i$ and $F_\mathrm{M}^{i+1}$  have collided if their frequency spacing $\lambda_\mathrm{M}^i=(F_\mathrm{M}^{i+1}-F_\mathrm{M}^i)/F_\mathrm{M}^i$ is less than the required spacing, $\lambda_\mathrm{min}$, i.e. $\lambda_\mathrm{M}^i<\lambda_\mathrm{min}$. The index $i=0, 1, \dots,N-1$ is used for the frequency order of $F_\mathrm{M}$ which may have changed with respect to $F_\mathrm{D}$, i.e. resonances may have swapped. $\lambda_\mathrm{min}$ is determined by the maximum level of accepted signal mixing between neighboring resonators, which strongly depends on the application and detector performance.  
The yield of the array $P_0$ is given as \cite{liuSuperconductingMicroresonatorArrays2017} 
\begin{equation}
P_0 =
\Bigg\{\prod_{n=1}^{n=N-1}
\Bigg[
1 -\frac{\mathrm{erf}\left(\frac{n \lambda_\mathrm{D} + \lambda_{\min}}{\sqrt{2}\sigma}\right)
-\mathrm{erf}\left(\frac{n \lambda_\mathrm{D} - \lambda_{\min}}{\sqrt{2}\sigma}\right)}{2}
\Bigg]\Bigg\}^2
\label{eq:yield}
\end{equation}
with $\mathrm{erf}$ the error function
\begin{equation}
    \mathrm{erf}(z) = \frac{2}{\sqrt{\pi}}\int_{0}^z e^{-t^2} dt
    \label{eq:error function}
\end{equation}
and depends on a choice for $\lambda_\mathrm{min}$, the designed frequency spacing $\lambda_\mathrm{D}$ and frequency scatter $\sigma$. The frequency scatter assumes that the measured resonances $F_\mathrm{M}[n]$ are normally distributed around their designed values $F_\mathrm{D}[n]$ with a standard deviation of $\sigma\times F_\mathrm{D}[n]$. Often, there is a frequency dependent trend in the measured frequency deviations $\epsilon_\mathrm{M}[n] = (F_\mathrm{M}[n]-F_\mathrm{D}[n])/F_\mathrm{D}[n]$ such that these are not normally distributed and its mean $\mu_\mathrm{M}\neq0$ and $\sigma_\mathrm{M}\neq\sigma$. The trend in frequency has little influence on the number of collisions as neighboring resonators get approximately the same overall shift. We can therefore fit and subtract the frequency trend to find the valid value for $\sigma$.

\begin{figure}[ht!]
    \centering
    \includegraphics[width=\linewidth]{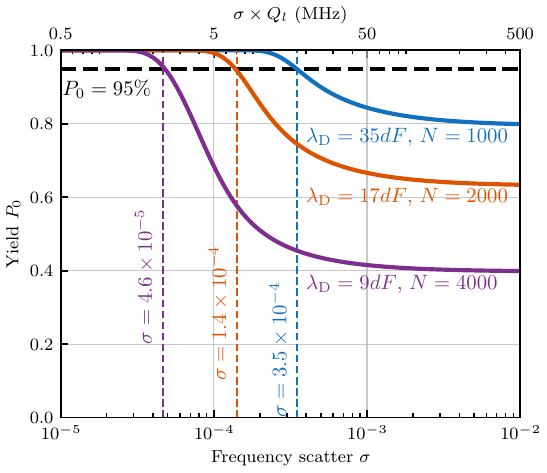}
    \caption{Yield $P_0$ as a function of frequency scatter $\sigma$ for $N$ detectors multiplexed on a single octave of readout bandwidth with a spacing $\lambda_\mathrm{D}$. The minimum required spacing is $\lambda_\mathrm{min}=4dF=\num{8.0e-5}$. The vertical dashed lines, matching in color to the values of $N$, indicate the required $\sigma$ to have $P_0=\qty{95}{\percent}$ (black dashed line).}
    \label{fig:multiplexing}
\end{figure}

To understand the impact of frequency scatter on the detector yield we plot, in \cref{fig:multiplexing}, a calculation of $P_0$ as function of $\sigma$ using \cref{eq:yield}. As an example, we set $\lambda_\mathrm{min}=4 dF= \num{8.0e-5}$, with $dF = F_\mathrm{D}/Q_l$ the resonator's linewidth defined by the loaded quality factor $Q_l=\num{5e4}$. For $P_0\geq\qty{95}{\percent}$ (dashed black line), we need $\sigma<\num{3.5e-4}$ for $N=\num{1000}$ (blue), $\sigma<\num{1.4e-4}$ for $N=\num{2000}$ (orange) and $\sigma<\num{0.5e-4}$ for $N=\num{4000}$ (purple). 
Typical values for frequency scatter in the literature are $\sigma=\num{1e-3}$ \cite{reyesAMKIDLargeKIDbased2025} to $\sigma=\num{1e-2}$ \cite{shuUnderstandingMinimizingResonance2021, middletonCCATLEDMapping2024, albertSpatialMappingKilopixel2024}. 
The lowest value for $\sigma$ reported without post-fabrication correction of the resonances is $\sigma=\num{4.7e-4}$ \cite{karatsuDESHIMA202004002026}. It is unknown how such low values for frequency scatter could be obtained, but it seems that the detector design used is more resilient to fabrication-induced variations as we use the same electron-beam fabrication process for our arrays.
Only  post-fabrication trimming of the resonators has led to sufficiently low values of $\sigma=\num{3.5e-4}$ \cite{liuSuperconductingMicroresonatorArrays2017, shuUnderstandingMinimizingResonance2021} and $\sigma=\num{1.8e-4}$ \cite{shuIncreasedMultiplexingSuperconducting2018}. 
% It is also possible to improve the resonator spacing \emph{in situ} without post-fabrication trimming by individually modifying the current bias to every resonator and using the nonlinear kinetic inductance to shift the resonance \cite{vissersSituFrequencyTuning2026}. The clever use of persistent current loops can reduce the number of bias lines required but this method is only applicable for MKIDs made of superconductors with a high critical temperature of around \qty{15}{K} and is therefore not applicable for single-photon-counting MKIDs that require superconductors with critical temperatures of around \qty{1}{K}.  
% There are also methods in development that can do active resonator control by modulating the applied readout drive current \cite{roubleFirstDemonstrationActive2024}. This method can be used to improve the resonator spacing and prevent collisions under optical load, but cannot be used to separate overlapping resonators. 

\begin{figure*}[ht!]
    \centering
    \includegraphics[width=\linewidth]{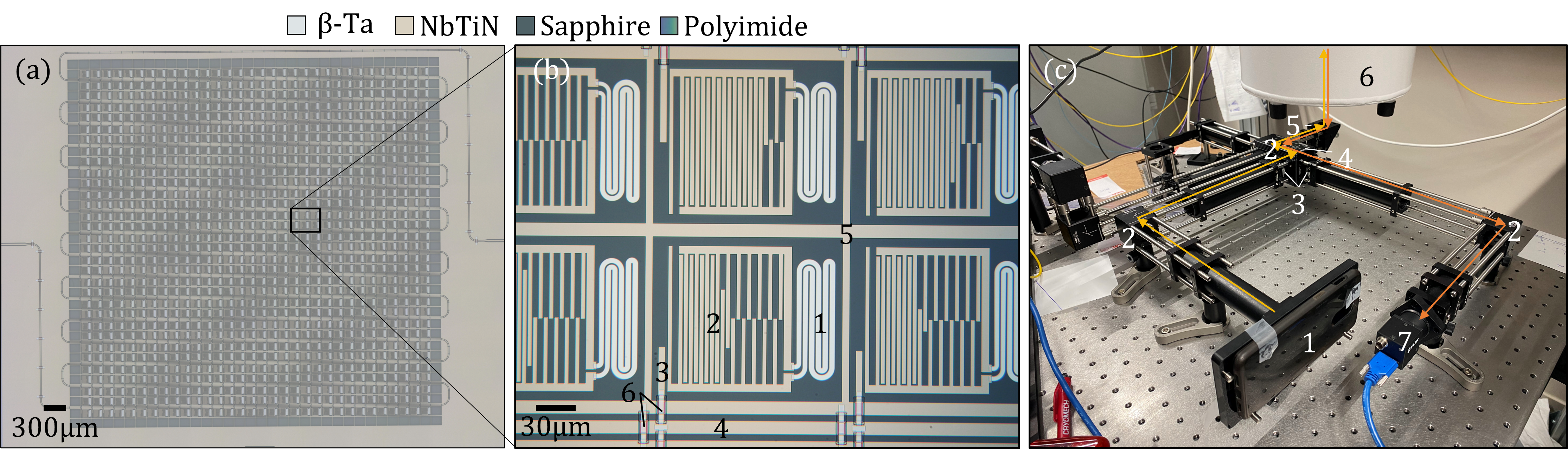}
    \caption{(a, b) Photographs of device A that is described in the main text. Indicated are: 1. inductor, 2. IDC, 3. coupler bar, 4. CPW center line, 5. ground plane, 6. grounding and coupling bridges. Further details on design and fabrication are found in the main text. (c) The pixel-to-frequency mapping setup with indicated components: 1. smartphone, 2. right angle mirrors, 3. \qty{300}{mm} imaging lenses, 4. $50:50$ beam splitter, 5. steering mirror, 6. cryostat, 7. monochrome camera. The yellow arrows show the optical path from phone to the device. The orange arrows show the the optical path from the device to the camera.}
    \label{fig:device and setup}
\end{figure*}

\subsection{Detector and array design}
In this work we present the results from three different devices: A, B, and C. The general detector and array design in this work is based on the design by \textcite{kouwenhovenResolvingPowerVisibleToNearInfrared2023}. Device A is our main device on which we perform the trimming, B and C are test devices to study the effect of the pixel pitch. We will give a detailed description of device A in the following and then briefly mention the differences of devices B and C.

\subsubsection{Device A} 
\label{sec:device A}
Photographs of device A are shown in \cref{fig:device and setup}a and b. 
The MKIDs have a hybrid design with a \ce{\beta-Ta} meandering inductor and an \ce{NbTiN} interdigitated capacitor (IDC) connected in parallel. The inductor is the light sensitive part where Cooper pairs are broken by the absorbed photons. Normally, a microlens array is glued on top of the detector array to focus the light onto the inductor \cite{kouwenhovenResolvingPowerVisibleToNearInfrared2023}. The meandering inductors are almost identical for all MKIDs. They all have an average width of \qty{4}{\um} and \qty{1}{\um} gaps, but are tapered differently depending the the detector's resonance frequency. The tapering gives a more uniform current distribution over the inductor and a more consistent single photon response \cite{mazinSuperconductingFocalPlane2012}. 
The IDC is used to set the resonance frequency of the detector \cite{igrejaAnalyticalEvaluationInterdigital2004}. The IDC has \num{10} finger pairs that are \qty{98}{\um} long and \qty{3}{\um} wide with a \qty{1}{um} gap. The fingers pairs are cut one by one to have the physical length necessary to define the 1024 resonance frequencies. They are also cut from both sides symmetrically to be more robust against misalignment in the lithography process. A minimum overlap of the fingers of \qty{1}{\um} is kept to reduce jumps in frequency. We parameterize the total overlap of the IDC finger pairs with $l_\mathrm{C}$ which ranges from \qtyrange{10}{970}{\um}. 
The IDC has a separate coupler finger that capacitively couples to a coupler bar. The coupler bar galvanically connects to the center line of the coplanar waveguide (CPW) readout line with a \ce{\beta-Ta} bridge suspended from ground by a \qty{600}{nm} thick polyimide patch. The length of the coupler bar is varied such that all detectors are overcoupled at $Q_l\approx Q_c=\num{5e4}$.
The length of the coupler bar and IDC fingers are determined from detailed simulations in Sonnet \cite{SonnetSoftware}. We find the resonance frequency and quality factor for the range of $l_\mathrm{C}$ with \qty{1}{\um} increments and at every value of $l_\mathrm{C}$ we simulate a range of coupler bar lengths. The design frequency and quality factors are then obtained through interpolation. We use the three port method to reduce the simulation time \cite{wisbeyNewMethodDetermining2014}.  
The \ce{NbTiN} is \qty{125}{nm} thick and has a critical temperature $T_c$ of \qty{11}{K} and sheet resistance $R_s$ of \qty{35}{\Omega}. The \ce{\beta-Ta} is \qty{41}{nm} thick, with $T_c=\qty{0.95}{K}$, $R_s=\qty{48}{\Omega}$ resulting in a sheet kinetic inductance $L_k=\qty{71}{pH\per\Box}$. 

The array has $\num{32}\times\num{32}$ pixels with a \qty{150}{\um} pixel pitch. The CPW meanders through the array and connects to detectors on both sides, one flipped with respect to the other. We place ground plane bridges every \qty{150}{\um}. We added holes in the ground plane surrounding the array to recreate the ground plane present within the array. The desired frequency range is from \qtyrange{4.15}{7.85}{GHz} with a \qty{300}{MHz} gap around \qty{6}{GHz} for the local oscillator (LO). An equal amount of \num{512} detectors are allocated on both sides of the LO-gap such that $\lambda_\mathrm{D}=34/Q_l=\num{6.7e-4}$ below the LO-gap and $\lambda_\mathrm{D}=24/Q_l=\num{4.8e-4}$ above the LO-gap. For the initial fabrication we design the resonance frequencies \qty{3}{\percent} lower than the desired frequency range such that we can trim the frequencies to the desired range in post-fabrication.

Resonators close in frequency should be placed far apart spatially to reduce electromagnetic cross-coupling \cite{noroozianCrosstalkReductionSuperconducting2012}. Detector $n=0$ with the lowest frequency is placed at the lower left corner. The frequencies on the same row are spaced as far apart in frequency as possible such that swapping of the resonances is highly unlikely. i.e. detectors $n=0, 32, \dots, 991$. We do this for the practical reason of speeding up the pixel-to-frequency mapping as we now only have to scan the individual rows and not the columns. Every consecutive row above skips \num{4} frequency positions and is also rolled horizontally with \num{10} frequency positions to increase the physical distance between pixels with the closest frequencies. So, the second row has detectors $n=4, 36, \dots, 996$, the third $n=8, 40, \dots, 1000$, continuing as such to the eighth row with $n=28, 60, \dots, 1020$. From the ninth row the pattern is repeated starting with $n=1, 33, \dots, 993$. This frequency allocation places the closest spatial neighbors more than \qty{2.7}{mm} apart and the closest frequency neighbors at least \qty{88}{MHz} apart.

\subsubsection{Devices B and C} 
\label{sec:devices B and C}
Devices B and C are identical to one another except for their pixel pitch. The fabrication of both devices is also identical as both devices had a similar off-center position on the same wafer. Device B is a \num{32}$\times$\num{32} array with pixel pitch of \qty{150}{\um}, just like device A. Device C is a copy of device B except that only every third row is kept and only every third detector in that row. This enlarges the pixel pitch to \qty{450}{\um}. 

The detector and array design of devices B and C differ slightly from device A. The \ce{\beta-Ta} layer of devices B and C is \qty{47}{nm} thick with $L_k=$\qty{56}{pH\per\Box}. The meandering inductors are not tapered but are \qty{4}{\um} wide with \qty{2}{\um} gaps. The inductor also has square instead of round corners. The capacitor fingers are \qty{4}{\um} wide and are cut from one side only. Devices B and C do not have the holes in the ground plane surrounding the array. The frequency range, spacing and allocation are approximately the same.
A photo of devices B and C is presented together with their measurement results in \cref{fig:sparse_vs_compact}.

\subsection{Fabrication}
We start with a $\varnothing \qty{100}{mm}$, \qty{350}{\um} thick, C-plane Sapphire wafer as our substrate. We deposit the \ce{NbTiN} using reactive magnetron sputtering in the LLS801 with the target in shuttling mode, known to produce highly uniform layer properties \cite{thoenSuperconductingNbTinThin2017}.
We pattern and etch the alignment markers for the subsequent layers with ultraviolet (UV) contact lithography and reactive ion etching (RIE) using \ce{SF_6} and \ce{O_2}. We clean the surfaces with an oxygen plasma using RIE before continuing to pattern the CPW, coupler bars and capacitors in the \ce{NbTiN} with electron beam (EB) lithography. We set the main fields to be $\qty{300}{\um}\times\qty{300}{\um}$ in size---twice the pixel pitch---and make sure the edges of the fields are always located in the ground plane to prevent stitching errors \cite{scholtenhuisEnhancedElectronbeamLithography2026}.  
We spin a positive resist CSAR ARP6200-13 at \qty{1300}{RPM} and bake it at \qty{150}{\celsius} for 3 minutes. To reduce charging of the resist we apply an extra layer of Electra E92 which we spin at \qty{2000}{RPM} and bake at \qty{90}{\celsius} for 2 minutes. We use a \qty{10}{nm} beam step size and a dose \qty{1100}{\upmu C \per cm^2}. We apply a proximity effect correction that optimizes the local dose to account for the back scattering of electrons. We use RIE with \ce{SF_6} and \ce{O_2} to etch the exposed \ce{NbTiN}. The polyimide layer for the ground plane and coupling bridges is spun and patterned with contact lithography after which it is developed and cured. We do a cleaning with oxygen plasma and use \ce{HF} for hydrogen passivation of the surface.

The \ce{\beta-Ta} for the inductors and bridges is sputter-deposited and patterned using a combined UV and EB lithography process that was developed by \textcite{thoenCombinedUltravioletElectronbeam2022a}. We use UV for the large ground planes outside the array and EB for the inductors and bridges. We spin the negative resist ma-N1405 at \qty{1500}{RMP} and bake for 3 minutes at \qty{100}{\celsius} before the patterning with a beam step size of \qty{10}{nm} and a dose \qty{1100}{\upmu C \per cm^2}.

The trimming of the capacitor fingers is done using the same process as used for the initial patterning and etching of the capacitors. How the trimming lengths are determined is explained in \cref{sec:trimming}.

\begin{figure*}[ht!]
    \centering
    \includegraphics[width=\linewidth]{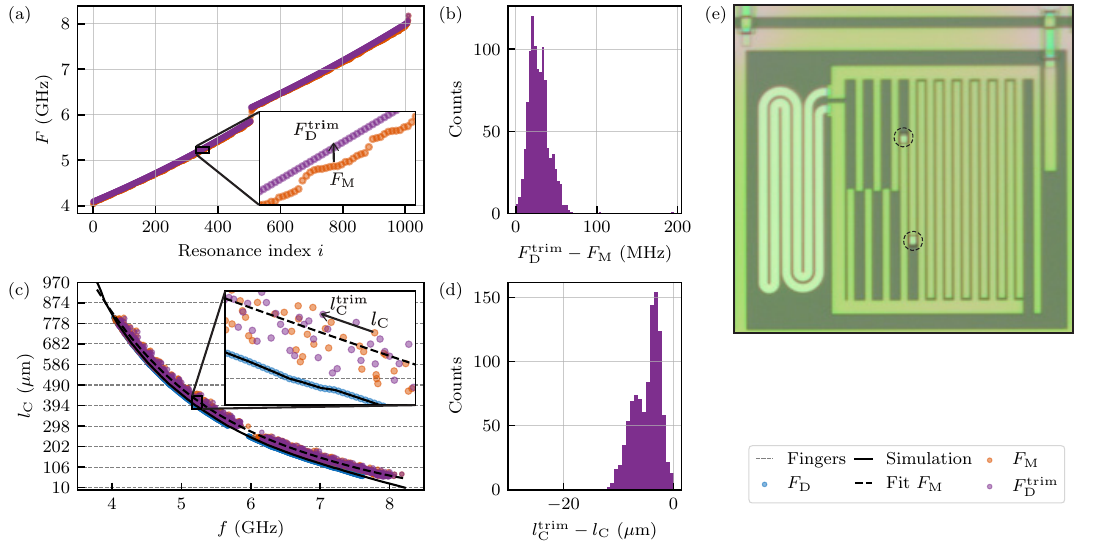}
    \caption{Overview of the post-fabrication trimming method used. (a) Based on the mapped resonances $F_\mathrm{M}$ (orange) a new set of design frequencies is determined with a constant frequency spacing $F_\mathrm{D}^\mathrm{trim}$ (purple). The frequency order of $F_\mathrm{M}$ is maintained to minimize the difference $F_\mathrm{D}^\mathrm{trim}-F_\mathrm{M}$. (b) Histogram of $F_\mathrm{D}^\mathrm{trim}-F_\mathrm{M}$. (c) Rather than using the simulated data (solid line), a 6-degree polynomial fit to $l_\mathrm{C}(F_\mathrm{M})$ (thick dashed line) is interpolated at $F_\mathrm{D}^\mathrm{trim}$ to find the corresponding finger lengths $l_\mathrm{C}^\mathrm{trim}$. The transition points between finger pairs are indicated by the horizontal dashed lines. (d) Histogram of $l_\mathrm{C}^\mathrm{trim}-l_\mathrm{C}$. (e) Photograph of a single pixel during the trimming process right after the patterning of the resist (green film). The lighter rectangles within the dashed circles are the exposed areas.}
    \label{fig:trimming}
\end{figure*}

\subsection{Measurement setup}
\label{sec:measurement setup}
The cryogenic setup consist of the dilution refrigerator described by \textcite{kouwenhovenResolvingPowerVisibleToNearInfrared2023}. We have two magnetic shields around the array, one made of superconducting Niobium and the other of Cryophy. These shields have long snouts specifically to block DC magnetic fields. The simulated shielding is of a factor \num{1e6} for time-dependent magnetic fields and a factor \num{70} for static fields \cite{RooijQuasiparticleDynamicsDisordered2026}. All the shields are mounted with transparent windows to illuminate the sample from outside of the cryostat. The device is typically operated at bath temperatures between \qtylist[list-units=single]{25; 100}{mK}. 

The measurements consist of complex $S_{21}$ transmissions with a vector network analyzer (VNA). The resonance frequencies are obtained as the minima in $S_{21}$ transmission. Crucial is to map the resonances to the spatial position of the pixel within the array. Our array is too compact to use a cryogenic LED-mapper as done in Refs. \cite{liuCryogenicLEDPixeltofrequency2017, martsenDevelopmentMKIDFrequencytopixel2025b, albertSpatialMappingKilopixel2024, middletonCCATLEDMapping2024}. Rather, we use an optical setup located outside the cryostat to illuminate the detectors row by row and column by column while tracking at which position of the lines the KID resonances respond, see \cref{fig:device and setup}c. This setup was inspired by Refs. \cite{walterMKIDExoplanetCamera2020, bottomSmartphoneSceneGenerator2018}. The line is generated using a smartphone, imaged by a single lens with $\qty{300}{mm}$ focal length and projected into the cryostat by a steering mirror. The setup has magnification of $-1$, with both the object and image at  $\qty{600}{mm}$ from the lens. The phone (OnePlus 11T) has an amoled screen and pixel pitch of \qty{50}{\um} enabling us to scan the line across the array with \qty{50}{\um} resolution. The lines are displayed by a custom-built Android application programmed using the Pygame module in Python. The measurement is automated by controlling the phone from the computer via Android Debug Bridge (ADB).
We added a $50:50$ beam splitter and a second lens that diverts half of the beam to a camera (BLKFLY-U3-50H5M) for visual alignment of the projected lines.

\subsection{Capacitor trimming}
\label{sec:trimming}
Capacitor trimming entails that we do a correction of the IDC finger lengths to improve the frequency spacing of the resonators. We can only shorten the fingers and increase the resonance frequencies. The trimming process has three steps that are explained in the following and visualized in \cref{fig:trimming}. 

First, we find a new set of design resonances, $F_\mathrm{D}^\mathrm{trim}$, see \cref{fig:trimming}a and b. We want to correct as little as possible by keeping the possibly swapped frequency order of the measured resonances $F_\mathrm{M}$. We also require that $F_\mathrm{D}^\mathrm{trim}$ has a constant fractional frequency spacing $\lambda_\mathrm{D}^\mathrm{trim}$. Lastly, we require that all resonators are trimmed to minimize the effect of a processing bias. An exception was made for three pixels with resonances relatively far above the rest of the array. These were not trimmed and allowed a larger frequency spacing. The frequency scatter after trimming is defined with respect to $F_\mathrm{D}^\mathrm{trim}$.

Second, we find the new finger lengths $l_\mathrm{C}^\mathrm{trim}$, see \cref{fig:trimming}c and d. We continue cutting the capacitor fingers exactly as was done in the original fabrication, see \cref{sec:device A}. This makes a separate trimming finger needless, reducing the detector footprint, and---more importantly---does not require any new simulations. We fit the measured dependency of $l_\mathrm{C}$ on $F_\mathrm{M}$ and interpolate at $F_\mathrm{D}^\mathrm{trim}$ to find $l_\mathrm{C}^\mathrm{trim}$. The fitting is done with a 6-degree polynomial. Panel d shows two distributions because of the slightly different $\lambda_\mathrm{D}^\mathrm{trim}$ above and below the LO-gap. The resonators that are not identified in the pixel-to-frequency mapping are all maximally trimmed to shift them to around \qty{9}{GHz} where they cannot cause frequency collisions. 

The third and final step of the trimming process is the fabrication of $l_\mathrm{C}^\mathrm{trim}$. This is done identical to the initial fabrication of the capacitors, see \cref{sec:device A}, except that we use a smaller beam step size of \qty{1}{nm}. In \cref{fig:trimming}e a photograph is shown of a single pixel right after the EB exposure. The green film is the resist and the lighter rectangles in the dashed circles are the patterned areas. The patches are \qty{4}{\um} wide and have a minium length of \qty{5}{\um}, extending over the substrate if the length to be trimmed is shorter. This is to make sure that even the shortest trimming lengths are properly developed and etched. Sapphire is not sensitive to over-etching so no significant trenches will be created in between the fingers.

% \clearpage

\section{Results}
\label{sec:results}
\begin{figure*}
        \centering
        \includegraphics[width=\linewidth]{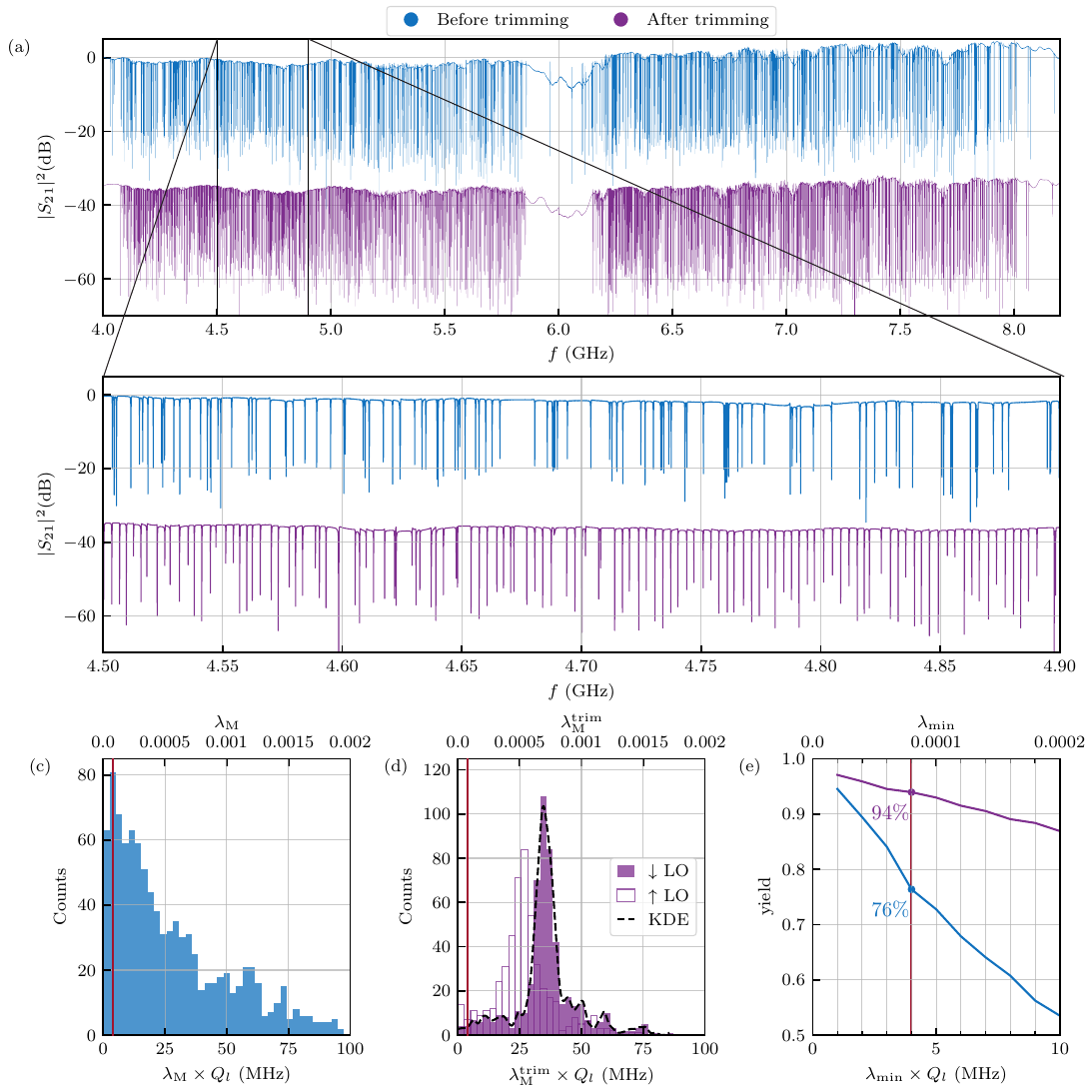}
        \caption{Yield from before and after post-fabrication trimming. Anything blue pertains to before trimming and anything purple to after trimming. (a) VNA transmission sweeps of the array from before and after trimming with an enlarged view of the indicated section. The gap around \qty{6}{GHz} is by design and reserved for the LO. (c) Frequency spacings of the resonators before trimming both in terms of resonator line widths ($dF=F_\mathrm{D}/Q_l$ with $Q_l=\num{50e3}$) and in dimensionless form. (d) Frequency spacings of the resonators after trimming with separate distributions for the upper and lower frequency bands around the LO-gap. A kernel density estimate (KDE, dashed line) applied to the lower band distribution reveals the standard deviation of the trimming method itself, $\sigma_\mathrm{lim}^\mathrm{trim}$. (e) Yield as a function of minimal required frequency spacing before trimming (blue) and after trimming (purple). The vertical red lines in panels c-e indicate the example value of $\lambda_\mathrm{min}=4 dF=\num{0.8e-4}$.}
        \label{fig:yield}
\end{figure*}

\subsection{Post-fabrication trimming}
\label{sec:results trimming}
We compare the yield of device A before and after trimming in \cref{fig:yield}, which is the main result of this work. The pixel-to-frequency mapping yielded 1012 out of 1024 resonators before trimming and 1008 after trimming. The drastic improvement in resonator spacing after trimming is clearly observed in the transmission data plotted in panel a.
We plot the distributions of the fractional frequency spacings $\lambda_M$ of the resonators before and after trimming in panels c and e, respectively. The distribution of $\lambda_M^{trim}$ consist of two groups because the detectors above the LO-gap are more tightly spaced than the ones below. The yield is obtained by setting a minimal required spacing $\lambda_\mathrm{min}$ indicated by the red vertical lines. The yield is equal to the fraction of detectors that are spaced further apart than $\lambda_\mathrm{min}$ from both their frequency neighbors. As the yield depends on a choice of $\lambda_\mathrm{min}$, we plot the yield $P_0$ for a range for $\lambda_\mathrm{min}$ in panel d. The graph shows that for $\lambda_\mathrm{min}=4dF=\num{0.8e-4}$ the yield improves from \qty{76}{\percent} to \qty{95}{\percent}. 

\begin{figure*}
        \centering
        \includegraphics[width=\linewidth]{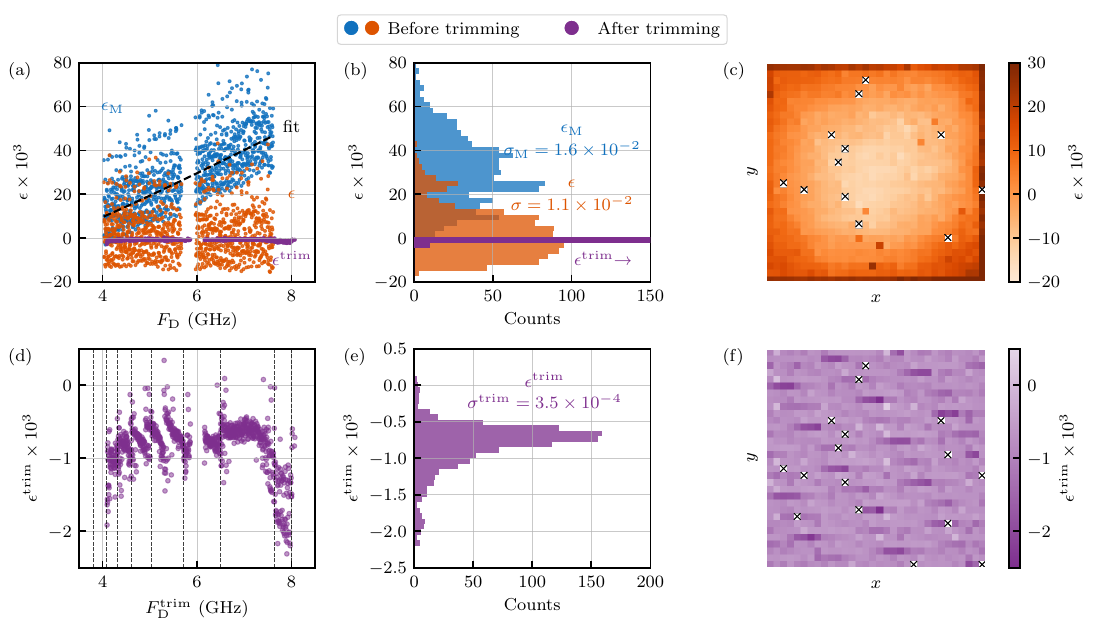}
        \caption{Frequency scatter from before and after post-fabrication trimming. (a) The frequency deviations before trimming $\epsilon$ (orange) are obtained from the measured frequency deviations $\epsilon_M$ (blue) by subtracting a quadratic fit (dashed line) and compared to the frequency deviations after trimming $\epsilon^\mathrm{trim}$ (purple). The values for the frequency scatter are annotated. No fit is subtracted after trimming. (b) Histograms of $\epsilon$, $\epsilon_M$ and $\epsilon^\mathrm{trim}$. The histogram $\epsilon^\mathrm{trim}$ is cut short for visibility, but continues to the right to around a thousand counts. (c) Two dimensional color map of $\epsilon$. Unidentified pixels are marked with a cross. (d,e) Enlarged view of $\epsilon^\mathrm{trim}$ from panels a and b. The vertical dashed lines in panel d mark the transitions to the cutting of a new capacitor finger pair. (f) Two dimensional color map of $\epsilon$$\epsilon^\mathrm{trim}$.}
        \label{fig:before_vs_after}
\end{figure*}

In \cref{fig:before_vs_after} we compare the frequency scatter from before and after trimming. In panel a and b we plot the measured frequency deviations $\epsilon_M$ before trimming (blue), from which we subtract a quadratic fit (dashed line) to obtain a gaussian-like distribution for the frequency deviations $\epsilon$ (orange). $\epsilon$ is not fully gaussian as it depends on the pixel position $(x, y)$ as seen in the spatial map plotted in panel c. These spatial effects are a property of our arrays caused by the pixel design in combination with the \qty{150}{\um} pixel pitch and are investigated further in \cref{sec:results pixel pitch}. 
We also plot the measured frequency deviations after trimming $\epsilon^\mathrm{trim}$ (purple) in panels a and b with an enlarged view in panels d and e. We show an improvement in the frequency scatter of over a factor 30 from $\sigma=\num{1.1e-2}$ before trimming to $\sigma^\mathrm{trim}=\num{3.5e-4}$ after trimming. No fit was subtracted after trimming, i.e. $\sigma^\mathrm{trim}=\sigma^\mathrm{trim}_{M}$. The spatial map of $\epsilon^\mathrm{trim}$ plotted in panel f also shows that we have corrected the spatial patterns observed in $\epsilon$. There is a striping pattern visible in panel f that is caused by the jumps in $\epsilon^\mathrm{trim}$ in panel d. These jumps occur when the cutting of the capacitor fingers transitions from one finger pair to the next (vertical dotted lines). An indication of these discrete steps between finger pairs is also visible in the simulation data, see the inset of \cref{fig:trimming}c. The fit to $F_M$ that was used to determine $F_\mathrm{D}^\mathrm{trim}$ has effectively smoothed over these jumps causing larger deviations at these transition points. Future trimming efforts should use a combination of the simulated and the measured data to try and correct for these transition points. We can estimate the lower limit of the frequency scatter for the trimming method presented in this work from the width of the distribution in \cref{fig:yield}d. The width is obtained with a kernel density estimate (KDE) and we then find the lower limit as $\sigma^\mathrm{trim}_\mathrm{lim}\approx\mathrm{std}(\lambda_M^\mathrm{trim})/\sqrt{2}=\num{4.9e-5}$. This value would enable multiplexing factors of nearly $4000$ per octave if $\lambda_\mathrm{min}=\num{8.0e-5}$, see \cref{fig:multiplexing}.
The key results for device A are summarized in \cref{tab:summary}.

The loaded quality factors after trimming $Q_l=\num{4.4\pm1.8e4}$ show no significant difference with the quality factors from before trimming $Q_l=\num{4.4\pm1.6e4}$.
The quality factors are obtained by a Lorentzian fit to the resonance curves of the detectors \cite{khalilAnalysisMethodAsymmetric2012, probstEfficientRobustAnalysis2015}. 
We also compare the average power spectral density of the noise from two random sets of \num{10} detectors from before and after trimming, each spanning the whole frequency range, and find no significant difference. The noise was taken as the time domain phase response of the detector sampled at \qty{1}{MHz} \cite{kouwenhovenResolvingPowerVisibleToNearInfrared2023}. 

\begin{table}[h!]
\begin{threeparttable}
\caption{Summary of key results from before and after the post-fabrication trimming of device A.}
\label{tab:summary}
\begin{tabular}{l S S S S[table-align-exponent = true]}
\toprule
\multicolumn{1}{l}{} &
  \multicolumn{1}{c}{\begin{tabular}[c]{@{}c@{}}Fabricated\tnote{1} \\ \end{tabular}} &
  \multicolumn{1}{c}{\begin{tabular}[c]{@{}c@{}}Mapped\tnote{1} \\ \end{tabular}} &
  \multicolumn{1}{c}{\textbf{\begin{tabular}[c]{@{}c@{}} Frequency \\scatter \end{tabular}}} & 
  \multicolumn{1}{c}{\begin{tabular}[c]{@{}c@{}}\textbf{Yield}\tnote{1,2}\\ \end{tabular}} \\ \midrule
\begin{tabular}[c]{@{}l@{}}Before \\ trimming\end{tabular} &
  \num{1023} &
  \num{1012} &
  \num{1.1e-2} &
  \num{782} 
  \\
\begin{tabular}[c]{@{}l@{}}After \\ trimming\end{tabular} &
  \num{1020} &
  \num{1008} &
  \num{3.5e-4} &
  \num{962} 
  \\ \bottomrule
\end{tabular}
\begin{tablenotes}
    \footnotesize
    \item [1] out of a total of 1024 pixels
    \item[2] using $\lambda_\mathrm{min}=\num{8.0e-5}$
\end{tablenotes}
\end{threeparttable}
\end{table}

\subsection{Pixel pitch}
\label{sec:results pixel pitch}

\begin{figure*}[ht!]
        \centering
        \includegraphics[width=\linewidth]{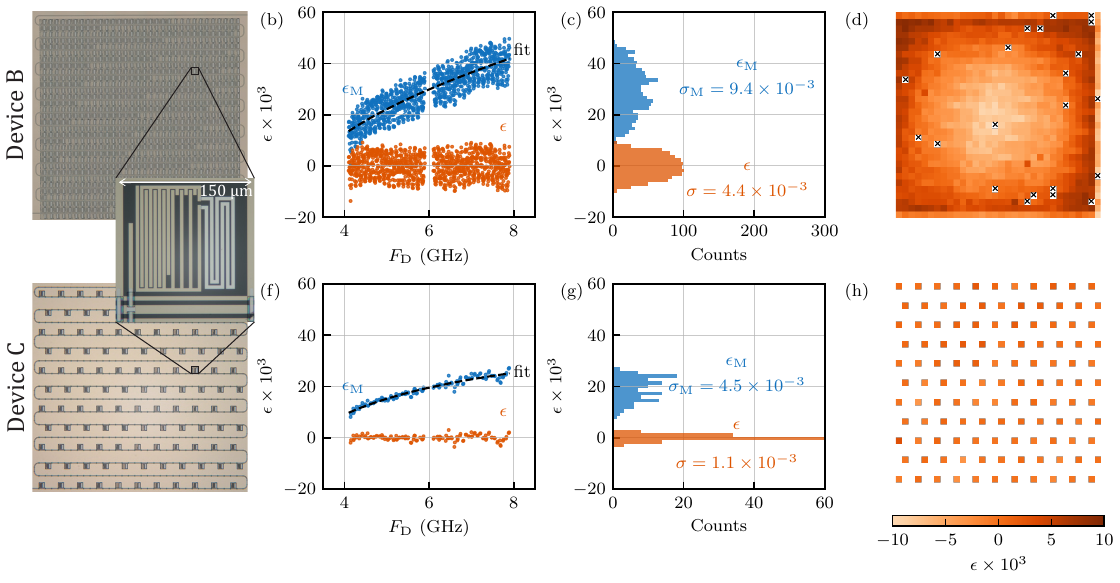}
        \caption{Frequency scatter before trimming impacted by the pixel pitch. (a-d) Results from device B with pixel pitch \qty{150}{\um} and (e-h) from device C with pixel pitch \qty{450}{\um}. Devices B and C are identical except for their pixel pitch, see also \cref{sec:devices B and C}. (a,e) Photographs of both devices and an enlarged view of one of their identical pixels. (b,f) Frequency deviations $\epsilon$ (orange) obtained from the measured frequency deviations $\epsilon_M$ (blue) by subtracting a quadratic fit (dashed line). (c, g) Histograms of $\epsilon_M$ and $\epsilon$ with their standard deviations annotated. (d, h) Two dimensional color map of $\epsilon$. Unidentified pixels are marked with a cross.}
        \label{fig:sparse_vs_compact}
\end{figure*}

We investigate the effect of the pixel pitch on the frequency scatter in \cref{fig:sparse_vs_compact}. We compare the results for device B in panels a-d and to those of device C in panels e-h. These devices are identical except that device B has a pixel pitch of \qty{150}{\um} and device C \qty{450}{\um}, see photographs in panels a and e, respectively. The shared inset highlights an identical detector in both arrays. The frequency deviations $\epsilon$ (orange) are obtained from the measured frequency deviations $\epsilon_M$ (blue) by subtracting a quadratic fit (dashed line), see panels b and f. Panels c and g show the distributions of $\epsilon$ and $\epsilon_M$ with their standard deviations indicated. Panels d and h show the spatial map of $\epsilon$ for both devices. Unidentified pixels are marked with a cross. The fabrication yield is \qty{98}{\percent} for device B and \qty{100}{\percent} for device C.

We find a factor \num{4} difference in $\sigma$ for devices B and C: \num{4.4e-3} and \num{1.1e-3}, respectively. This difference can only be caused by the difference in pixel pitch as the pixel design and fabrication as well as measurement and analysis are exactly the same for both devices. 
The spatial maps of $\epsilon$ reveal why the frequency scatter is worse for device B (panel d) compared to device C (panel h). Clear spatial patterns in $\epsilon$ are observed for device B, but not for device C. These spatial patterns increase $\sigma$ and increase the odds of frequency collisions because frequency neighbors are located far apart spatially. 
So, $\sigma=\num{1.1e-3}$ is the lowest value that we can obtain for this pixel design and fabrication method which shows the clear need for a post-fabrication correction method.

We identify two distinct spatial patterns in panel d: (1) a radially symmetric increase in $\epsilon$ from the center of the array towards the edges and (2) discrete offsets in $\epsilon$ at the edges at the top, bottom and right edges. These spatial patterns must have different causes as they are counteracting and have different spatial correlation lengths.
The radial pattern for the inner pixels is likely related to the electromagnetic (EM) environment of the detector that gradually changes across the array. This is not caused by EM cross-coupling of the pixels as this was not observed in the response of the detectors during the pixel-to-frequency mapping.
The offsets at the edges are likely caused by a sudden difference in the magnetic environment experienced by the pixels over a distance of no more than a pixel pitch. This is because the inductor is always on the right side of the pixels while the left edge is the one unaffected. Moreover, the top and bottom edges have an equal offset because they are symmetric, but the offset is less than for the right edge.
Simulations with up to $3\times2$ pixels in Sonnet could not reproduce these spatial effects.  
It is unlikely to be caused by magnetic flux trapping in the \ce{NbTiN} ground plane  outside the array \cite{baiFluxTrappingNbTiN2026} as the holes in the ground plane around the array of device A caused the same offsets but with opposite sign, see \cref{fig:device and setup}a and \cref{fig:before_vs_after}c. It is also not flux trapping in the \ce{\beta-Ta} inductors because the left and right edges are identical in that regard. 

The dependence of $\epsilon_M$ on $F_\mathrm{D}$ is present in both devices and not impacted by the difference in pixel pitch. Therefore, it must be intrinsic to the resonator design rather than the array design. It must also be caused by the capacitor as this is the only component that changes with $F_\mathrm{D}$, perhaps by a change in overall linewidth of the capacitor fingers or an over-etch in the substrate.

% \clearpage

\section{Conclusions}
\label{sec:conclusions}
In this work we demonstrate the post-fabrication trimming of a 1024-pixel MKID array with a pixel pitch of \qty{150}{\um}. We show an increase in yield from \qty{76}{\percent} before trimming to \qty{94}{\percent} after trimming. This increase in yield is caused by the reduction in frequency scatter of over a factor 30. The trimming method developed here is easily scalable to larger arrays and has no effect on the noise or the quality factor of the resonators. We discuss ways of how this method can be further improved.

Additionally, we show that---before trimming---the \qty{150}{\um} pixel pitch increases the frequency scatter with a factor 4 compared to an array with a pixel pitch of \qty{450}{\um}. The increase in frequency scatter is caused by spatial patterns observed across the array which are now under further investigation. Nonetheless, we show the effect of the pixel pitch on the frequency scatter is static and can be corrected in trimming. 

The raw data and code are made available on Zenodo to reproduce all the results in this work: \url{https://doi.org/10.5281/zenodo.22675240}.

% \clearpage

\begin{acknowledgments}
    This work is financially supported by the Netherlands Organisation for Scientific Research NWO (Vidi 213.149). We would like to thank Nick de Keijzer for designing adjustments to the setup for this experiments, and the SRON workshop for producing those. We are grateful to Nathan Bhoedjang, Gabriël Lomans and Maxim Herman for contributing to the measurement setup and analysis in their research projects.
\end{acknowledgments}

% \clearpage

\appendix

% \section{Interdigitated capacitance}
% \label{app:IDC}
% \input{SEC_app_IDC}

% \section{Pixel-to-frequency mapping}
% \label{app:mapping}
% \input{SEC_app_mapping}

% \section{Frequency allocation}
% \label{app:freq_alloc}
% \input{SEC_app_freq_alloc}

% \section{Resonator spacing}
% \label{app:resonator_spacing}
% \input{SEC_app_resonator_spacing}

% \section{Spatial effects}
% \label{app:spatials}
% \input{SEC_app_spatials}

% \section{Device A}
% \label{app:devA}
% \input{SEC_app_trimA}

\bibliography{bibliography}

\end{document}